# Experimental characterization of an ultra-broadband dual-mode symmetric Y–junction based on metamaterial waveguides

Raquel Fernández de Cabo[a,*], Jaime Vilas[a,b], Pavel Cheben[c], Aitor V. Velasco[a], David González-Andrade[d]

[a] *Instituto de Óptica Daza de Valdés, Consejo Superior de Investigaciones Científicas (CSIC), 121 Serrano, Madrid 28006, Spain*
[b] *Alcyon Photonics S.L., 11 Génova, Madrid 28004, Spain*
[c] *National Research Council Canada, 1200 Montreal Road, Bldg. M50, Ottawa K1A 0R6, Canada*
[d] *Centre de Nanosciences et de Nanotechnologies, CNRS, Université Paris-Saclay, Palaiseau 91120, France*
* Corresponding author: r.fernandez@csic.es

**ABSTRACT**

Silicon photonic integrated circuits routinely require 3-dB optical power dividers with minimal losses, small footprints, ultra-wide bandwidths, and relaxed manufacturing tolerances to distribute light across the chip and as a key building block to form more complex devices. Symmetric Y–junctions stand out among other power splitting devices owing to their wavelength-independent response and a straightforward design. Yet, the limited resolution of current fabrication methods results in a minimum feature size (MFS) at the tip between the two Y–junction arms that leads to significant losses for the fundamental mode. Here we propose to circumvent this limitation by leveraging subwavelength metamaterials in a new type of ultra-broadband and fabrication-tolerant Y–junction. An exhaustive experimental study over a 260 nm bandwidth (1420–1680 nm) shows excess loss below 0.3 dB for the fundamental transverse-electric mode ($TE_0$) for a high-resolution lithographic process (MFS ~ 50 nm) and less than 0.5 dB for a fabrication resolution of 100 nm. Subwavelength Y–junctions with deterministically induced errors of ±10 nm further demonstrated robust fabrication tolerances. Moreover, the splitter exhibits excess loss lower than 1 dB for the first-order transverse-electric mode ($TE_1$) within a 100 nm bandwidth (1475–1575 nm), using high-resolution lithography.

## 1. Introduction

Photonic integrated circuits (PICs) built on the silicon-on-insulator (SOI) platform benefit from high modal confinement, small footprints, energy efficiency and large-scale production, whilst driving-down costs thanks to the compatibility with complementary metal-oxide-semiconductor (CMOS) fabrication processes [1–3]. These compelling advantages substantially extend the scope of photonic integration beyond telecom and datacom to emerging applications with a far-reaching impact. These include 5G mobile communications [4], the Internet of Things [5], quantum photonics [6], light detection and ranging [7], spectrometry [8,9] and biochemical sensing [10], also enabling lab-on-a-chip solutions [11].

The complexity leap of the aforementioned applications requires an increasing number of on-chip components that take advantage of either multimode or broadband operations. Specifically, 3-dB optical power splitters are key components extensively used in light distribution or as building blocks for more intricate arrangements, including optical switches, multiplexers or integrated Mach-Zehnder interferometers [12,13].

Sequentially concatenated 3-dB power splitters are often utilized to implement 1×N dividers [14,15], requiring compact and low-loss designs. For datacom applications such as mode-division multiplexing [16] or multitarget sensing [17], power splitters with broad bandwidths are required. Different power division architectures have been reported based, among others, on symmetric Y–junctions [18,19], multimode interference (MMI) couplers [20–22], inverse tapers [23], adiabatic tapers [24], directional and adiabatic couplers [25–27], slot waveguides [28] and photonic crystal structures [29–31]. MMI devices offer good fabrication tolerances and compact footprints, and their operational bandwidth can be optimized through geometry design [20,21] or partially shallowly etched regions [22]. Inverse tapers [23] provide efficient mode evolution but typically present narrow bandwidths. Adiabatic tapers [24] exhibit wider bandwidths, but the performance of transverse-electric (TE) polarized light has larger fluctuations across the operation bandwidth due to the higher sensitivity to sidewall roughness. Conversely, the bandwidth of conventional directional couplers is restricted due to the strong wavelength-dependence of the evanescent coupling. Asymmetric [25] and bent [26] directional couplers have been demonstrated with reduced wavelength dependence, but it is not sufficient for ultra-broadband optical systems. Additional power division architectures with low losses over extensive operational bandwidth also include adiabatic couplers [27], slot waveguides [28] photonic crystal power splitters [29–31], inverse design methods [32,33], pixelated metastructures [34], and deep neural networks [35].

Amongst these power splitters, symmetric Y–junctions are one of the most common alternatives given their polarization- and wavelength-independent response, and their straightforward design. Symmetric Y–junctions consist of a stem waveguide ramifying into two arms of the same width. However, these structures typically present significant loss for fundamental modes at the junction, especially as the tilt angle between the two branching arms increases [36]. Two basic mechanisms are responsible for this loss: the wavefront mismatch due to the abrupt tilt angle, and the transformation of the mode profile in the interface between the stem and the arms [37]. In order to reduce the effect of the tilt angle, s-bend shaped waveguides can be employed for the arms [38], whereas the junction region can be tapered to ensure an approximately adiabatic mode evolution. Despite these design optimizations, minimum feature size (MFS) limitations of fabrication technologies lead to an imperfect tip between the two Y–junction arms. This MFS constraint causes considerable losses on the fundamental mode as the maximum of its power profile coincides with the junction tip [39]. Conversely, this same effect is negligible for the first-order mode, which presents a minimum of its power profile at the waveguide center. Several approaches to optimized Y–junction designs have been proposed to circumvent the effect of the MFS at the tip, including tapered and slotted waveguides [19,27,28], core size optimization [40] or particle swarm optimization algorithms [18]. Nevertheless, ultra-broadband, low-loss and fabrication-tolerant solutions are still sought after.

Subwavelength gratings (SWG) provide a powerful tool for improving the performance of a wide range of photonic devices [41–45]. SWGs are based on periodic grating waveguides with a period ($\Lambda$) significantly smaller than the operating wavelength ($\lambda$), hence behaving as a homogenous metamaterial (i.e. a medium with tailored optical properties). This behavior inhibits diffraction and enables refractive index and dispersion engineering [44]. Subwavelength engineering has become a strong design method for the realization of integrated silicon photonics components, including fiber-chip couplers [41], reconfigurable filters [46] and gradient-index lenses [47]. SWG metamaterials have also been successfully applied to several power splitters such as asymmetric directional couplers [48], three-guide directional couplers [49], inverse tapers [45,50], slot adiabatic waveguides [51], MMI devices [52–54], waveguide crossings [55] and Y–junctions [56]. Recently, we proposed an architecture for a high-performance and fabrication-tolerant SWG Y–junction [57]. However, this previous study only covered simulation results and preliminary measurements for $TE_0$.

In this paper, we present a comprehensive experimental study of a dual-mode Y–junction engineered with subwavelength metamaterials for deconfinement of the fundamental mode near the junction tip and mitigating losses. Specifically, we greatly expand our original study [57] by including accurate $TE_0$ mode measurements through cascaded splitters and $TE_1$ mode measurements through auxiliary mode multiplexers, for both SWG and conventional Y-junctions. We also measure biased devices in order to study the fabrication tolerances of our device considering both 50 nm and 100 nm MFSs. Our power splitter yields a measured fundamental transverse-electric mode ($TE_0$) excess loss (EL) of less than 0.3 dB considering a high-resolution fabrication process with MFS of 50 nm and below 0.5 dB for a resolution scenario with MFS of 100 nm within the 1420–1680 nm wavelength range. Moreover, the splitter exhibits excess loss lower than 1 dB for the first-order transverse-electric mode ($TE_1$) within a 100 nm bandwidth (1475–1575 nm), in the high-resolution process.

## 2. Device design

Our proposed SWG Y–junction, presented in Fig. 1(a), operates according to similar principles as a conventional symmetric Y–junction, illustrated in Fig. 2(a). The structure of the SWG and conventional Y–junctions comprise two single-mode output arms and a multimode input stem waveguide that supports the first two TE modes. When $TE_0$ is injected into the stem waveguide, its power is equally divided into two in-phase $TE_0$ modes, one in each output arm. When the stem waveguide is excited with $TE_1$, two $TE_0$ modes of equal power but with a relative phase difference of $\pi$ are generated in the two output arms [39]. Therefore, notice that the Y-junction operates simultaneously as a splitter and a mode converter for $TE_1$ mode. In the conventional symmetric Y–junction, MFS limitations at the junction tip results in significant $TE_0$ loss penalty. By applying SWG engineering, modal confinement near the junction tip is reduced and the $TE_0$ mode transition at the stem-arm interface is smoothed, significantly reducing losses.

Both splitters were designed for a 220-nm-thick Si core. Our SWG Y–junction includes an input strip waveguide of width $W_s = 1.2$ μm, optimized to avoid a weak confinement of the Bloch–Floquet $TE_1$ mode that would lead to high $TE_1$ excess losses ($EL_{TE1}$) due to substrate leakage or mode radiation. Using an adiabatic input taper of length $L_{ti} = 10$ μm, the strip waveguide is progressively adapted to a subwavelength stem waveguide of the same width and a length $L_c = 13$ μm. The SWG stem splits into two SWG single-mode arms of width $W = 500$ nm and length $L_b = 12.3$ μm. Each SWG arm is shaped as an s-bend to avoid abrupt tilt angles at the junction tip. After the SWG s-bends, each arm waveguide transforms into a strip output waveguide by means of an output taper of length $L_{to} = 6$ μm. The final separation between arms ($H_a$) is 1.5 μm. An initial arm offset ($H_{off}$) is included to account for two alternative MFS scenarios. We considered an MFS of 50 nm for high-resolution fabrication processes and an MFS of 100 nm.

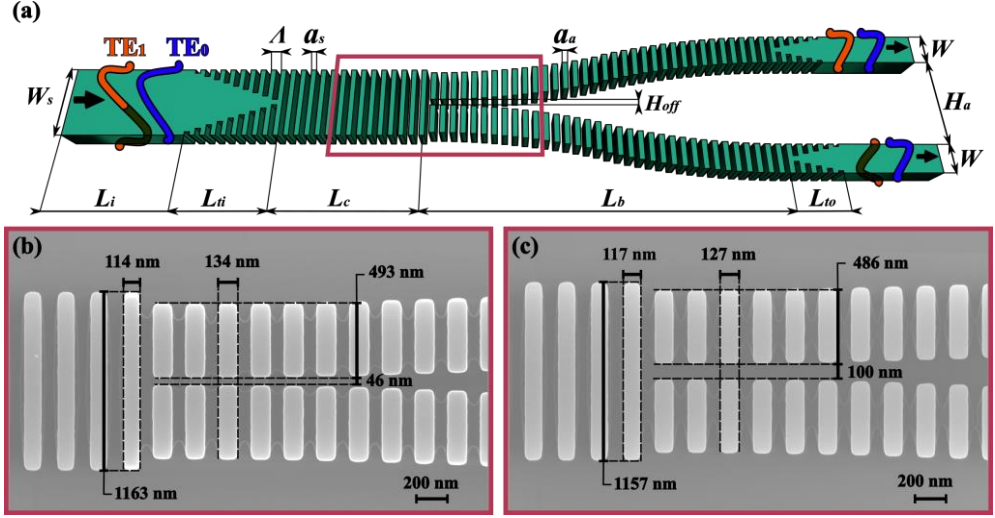

**Fig. 1.** Subwavelength Y–junction schematic (a) and SEM images for devices with an MFS of 50 nm (b) and 100 nm (c).

In both SWG arms and stem, $\Lambda$ was set to 220 nm to prevent Bragg reflection while maintaining SWG feature sizes above the target MFS. In order to minimize mode mismatch at the junction, we set different duty cycles in the stem ($DC_s = a_s/\Lambda$) and the arms ($DC_a = a_a/\Lambda$), where $a_s$ and $a_a$ are the lengths of the SWG silicon segments in the stem and arms, respectively. For $DC_s = 0.5$, the best EL balance between $TE_0$ and $TE_1$ modes was achieved for $DC_a = 0.6$ when considering an MFS of 50 nm and for $DC_a = 0.55$ with MFS of 100 nm. The design procedure followed is described in further detail in [57]. Notice that fabrication tolerances and the effect of temperature variations was also studied, showing negligible performance degradation for ±10 K. Figures 1(b) and 1(c) show scanning electron microscope (SEM) images of the fabricated SWG Y–junctions for MFS 50 nm and 100 nm, respectively.

In order to compare the performance of our SWG device with a conventional splitter, we designed conventional Y–junctions with different MFSs (Fig. 2(a)). These devices also comprise an input multimode stem waveguide of width $W_0 = 1$ μm, which is adapted to the width at the junction ($W_t = 2W + H_{off}$) by means of a taper of length $L_t = 4$ μm. The two single-mode s-shaped output arms maintain the same arm width ($W$), length ($L_b$) and final separation ($H_a$) as in the SWG splitter. The initial offset between the arms ($H_{off}$) was also included to consider the MFSs of 50 nm and 100 nm. Additionally, we also included a third study-case with $H_{off} = 0$ nm, which ideally would result in a perfect tip. SEM images of the fabricated conventional Y–junctions for $H_{off}$ of 0 nm, 50 nm and 100 nm can be seen in Figs. 2(b), 2(c) and 2(d), respectively. It can be observed that conventional Y–junctions with $H_{off} = 0$ nm and $H_{off} = 50$ nm, as fabricated, present comparable tip dimensions. This leads to almost analogous performance for both devices, as discussed in more detail in the following sections.

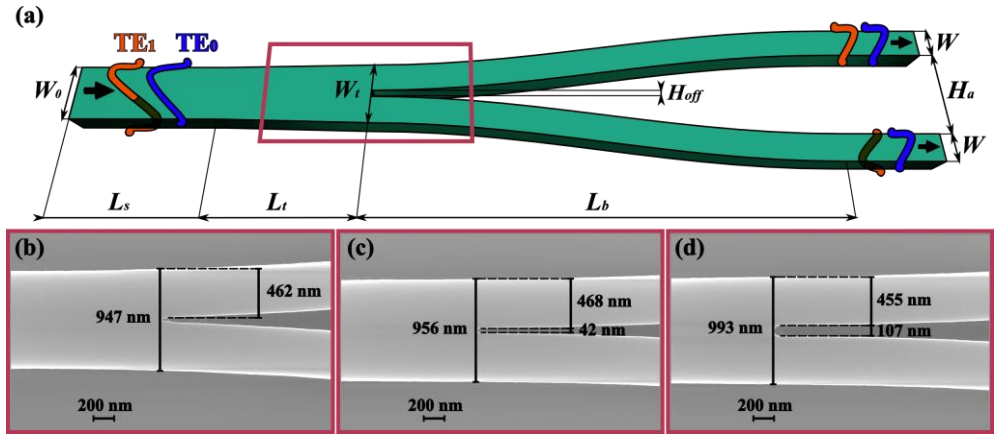

**Fig. 2.** Conventional Y–junction schematic (a), and SEM images of the tip for three resolutions: ideal 0 nm (b), and realistic 50 nm (c) and 100 nm (d).

## 3. Fabrication and experimental characterization

The device was fabricated using electron-beam lithography in a commercial foundry [58]. The SOI wafer has a silicon layer thickness of 220 nm and a 2-µm-thick buried oxide (BOX). The mask pattern was defined by exposing the resist to a 100 keV electron-beam lithography system, followed by an anisotropic reactive ion etching process that transfers the pattern to the Si layer. A SiO$_2$ cladding with a thickness of 2.2 µm was deposited by chemical vapor deposition. Finally, a deep etch process was applied to smooth the chip facets, allowing efficient fiber-chip edge coupling by using high-efficiency broadband SWG edge couplers [41].

Experimental characterization was carried out using two tunable lasers to sweep the wavelength from 1420 nm to 1680 nm, coupled to a three-paddle fiber polarization controller, a linear polarizer and a half-wave plate. TE polarized light was coupled into the chip using a lensed polarization-maintaining optical fiber. Light at the chip output was collected by a 40× microscope objective, directed to a Glan-Thompson polarizer, and focused onto a germanium photodetector.

*3.1 Fundamental transverse-electric mode (TE$_0$)*

In [57], 3D finite-difference time-domain simulations were performed for the SWG Y-junction, which provided excess loss for the TE$_0$ ($EL_{TE0}$) mode below 0.3 dB in a 350 nm bandwidth for the worst-case MFS scenario of 100 nm. Given the challenge of measuring losses of this order of magnitude in a stand-alone configuration, we implemented cascaded structures with multiple stages. That is, 1 to 4 stages of concatenated Y–junctions: 1×2, 1×4, 1×8 and 1×16 structures. SEM image of a 1×16 structure and the different stages are shown in Fig. 3(a), with a close-up view of the SWG Y–junction in Fig. 3(b). The 50 nm MFS is within the limit of the fabrication resolution offered by the foundry and, as it can be seen in the SEMs, the device was fabricated correctly. Reference waveguides with the same length and number of bends as the cascaded structures were also included to determine Y–junction excess loss.

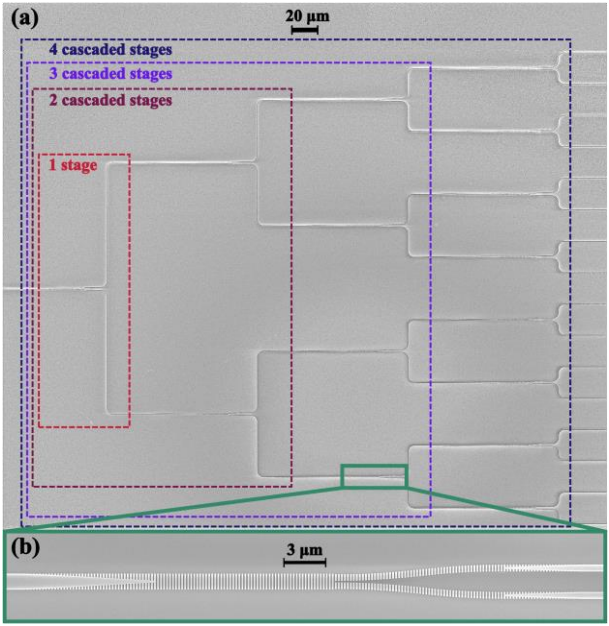

**Fig. 3.** SEM images of the SWG Y–junction in cascaded configuration 1×16 (a) and inset of the device (b).

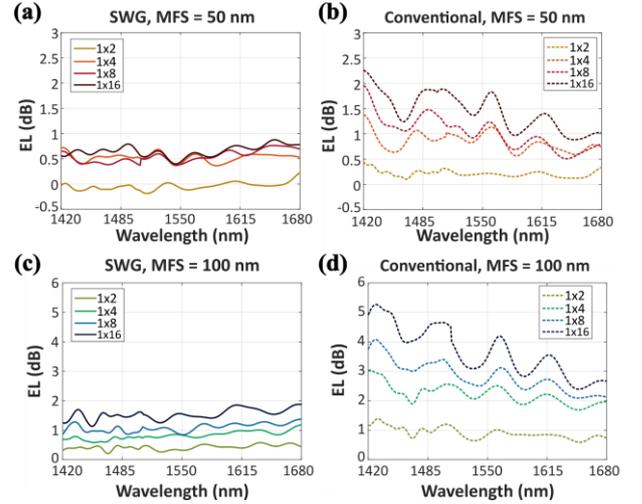

**Fig. 4**. Excess loss of the 1×2, 1×4, 1×8 and 1×16 cascaded structures, for SWG Y–junction with 50 nm MFS (a), conventional Y–junction with 50 nm MFS (b), SWG Y–junction with 100 nm MFS (c), and conventional Y–junction with 100 nm MFS (d).

The measured excess loss for different cascaded Y–junctions is presented in Fig. 4. The $EL_{TE0}$ of the SWG splitter designed for MFS = 50 nm ($EL_{TE0}^{SWG,50}$) is plotted in Fig. 4(a) and for MFS = 100 nm ($EL_{TE0}^{SWG,100}$) in Fig. 4(c). The $EL_{TE0}$ of the conventional Y–junction is shown in Fig. 4(b) for MFS = 50 nm ($EL_{TE0}^{Conv,50}$) and for MFS = 100 nm ($EL_{TE0}^{Conv,100}$) in Fig. 4(d). Conventional Y–junctions with $H_{off}$ = 0 nm were also measured but have not been included in Fig. 4 for clarity, since they are very similar to results with $H_{off}$ = 50 nm. This similarity is caused by the MFS limitation of the fabrication process and verifies our initial assumption on e-beam fabrication MFS. That is, even when the nominal design includes a perfect tip ($H_{off}$ = 0), experimental MFS limitations will induce an imperfect tip, in this case, similar to the design of $H_{off}$ = 50 nm. Notwithstanding, results for this $H_{off}$ are included in Figs. 5 and 8 for direct comparison. For both MFSs of 50 nm and 100 nm, the SWG splitter has a substantially reduced $EL_{TE0}$ compared to the conventional Y–junction.

Figure 4 shows the overall excess loss increment as more cascaded stages are included. To obtain the $EL_{TE0}$ relative to a single Y-junction, we performed a linear regression with the measured losses in each cascaded stage. Figure 5 shows the resulting $EL_{TE0}$ per splitter, that is, the average loss per cascaded splitter calculated through the slope of the linear regression. The SWG Y–junction (solid line) exhibits a flat response in the entire measured bandwidth from 1420 nm to 1680 nm, with an $EL_{TE0}^{SWG,50}$ lower than 0.3 dB and $EL_{TE0}^{SWG,100}$ below 0.5 dB in the full 260 nm spectrum. By contrast, the conventional Y–junction (dotted line) shows a performance degradation towards shorter wavelengths, resulting in a loss penalty, especially for an MFS of 100 nm (see Fig. 5(b)). Considering the high-resolution fabrication (MFS

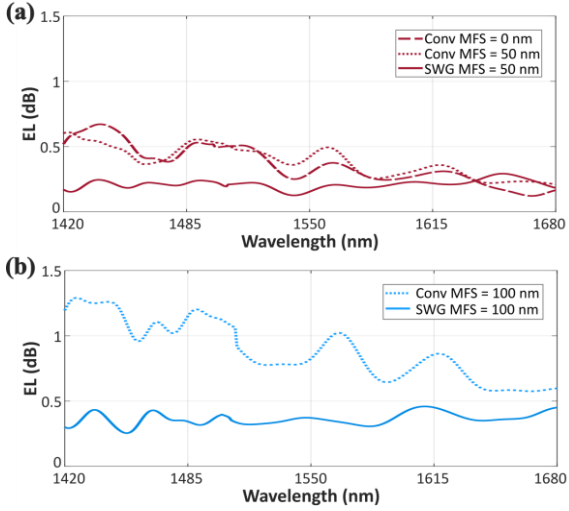
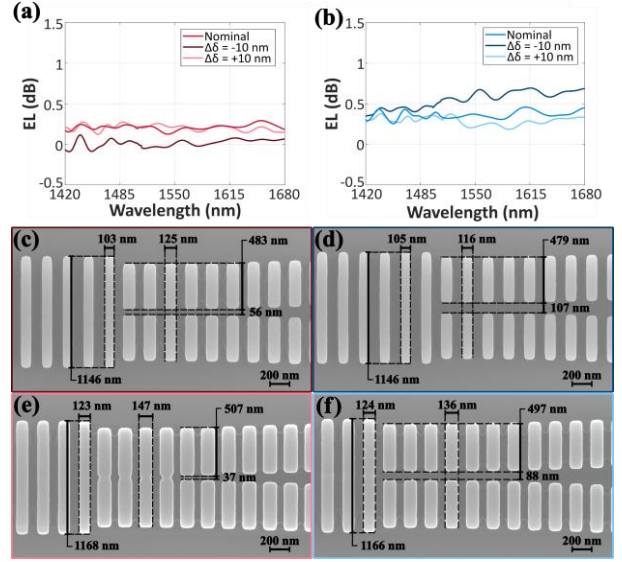

**Fig. 5.** EL per splitter measured through linear regression of the response of four cascaded stages for SWG (solid line) and conventional (dotted line) Y–junctions, for MFS of 50 nm (a) and 100 nm (b). Results obtained for the conventional splitter with ideal resolution (MFS = 0 nm) are also shown in panel (a).

**Fig. 6.** Tolerances to fabrication errors of $\Delta\delta = \pm 10$ nm for the SWG Y–junction with MFS of 50 nm (a) and 100 nm (b). Excess loss per splitter was measured through linear regression of cascaded stages. SEM images of devices with $\Delta\delta = -10$ nm for an MFS of 50 nm (c) and 100 nm (d). SEM images of devices with $\Delta\delta = +10$ nm for an MFS of 50 nm (e) and 100 nm (f).

= 50 nm), our device consistently outperforms the conventional Y–junction, with $EL_{TE0}^{SWG,50} < 0.23$ dB in a 215 nm bandwidth (1420 nm – 1635 nm) as presented in Fig. 5(a). As previously explained, it can be seen that excess losses for conventional Y-junctions with an $H_{off} = 0$ nm and $H_{off} = 50$ nm are very similar, only differing in a mean deviation of less than 0.06 dB. As the MFS increases to 100 nm, the negative impact on the performance of the conventional Y–junction is more pronounced, with an $EL_{TE0}^{Conv,100}$ above 0.57 dB in the 1420 nm –1680 nm window. The SWG device, on the other hand, yields improved performance over the entire measured bandwidth with $EL_{TE0}^{SWG,100}$ below 0.46 dB.

Devices with deterministically induced dimension variations ($\Delta\delta$) of $\pm 10$ nm were incorporated in the mask layout to measure the robustness of the SWG Y–junction to fabrication errors. Figure 6(a) shows $EL_{TE0}^{SWG,50}$, demonstrating that performance is preserved despite the presence of geometric variations, and even exhibiting a slight improvement for over-etching deviations (i.e., $\Delta\delta = -10$ nm, SEM shown in Fig. 6(c)). In contrast, for the MFS of 100 nm (Fig. 6(b)), $EL_{TE0}^{SWG,100}$ is slightly improved for $\Delta\delta = +10$ nm (SEM shown in Fig. 6(f)). For $\Delta\delta = -10$ nm (SEM shown in Fig. 6(d)), $EL_{TE0}^{SWG,100}$ performance degrades towards longer wavelengths. The largest fabrication bias was observed in waveguide width, narrowing the designed stem waveguide ($W_s = 1200 \pm 10$ nm) by approximately 40 nm.

### 3.2 First-order transverse-electric mode (TE$_1$)

In order to characterize the TE$_1$ mode division, a mode multiplexer [59] was included in combination with the SWG Y–junction, as schematically shown in Fig. 7. When TE$_0$ is injected through the upper input port of the mode multiplexer, the TE$_0$ mode is generated at the output multimode waveguide. When the lower input port is excited with TE$_0$, mode evolution results in TE$_1$ at the mode multiplexer output. Two mode multiplexers in back-to-back were used as reference to extract TE$_1$ mode excess loss of the Y–junctions.

Figure 8(a) shows $EL_{TE1}$ measurements for MFS = 50 nm in both SWG ($EL_{TE1}^{SWG,50}$) and conventional ($EL_{TE1}^{Conv,50}$) Y–junctions, as well as for the conventional Y–junction with $H_{off} = 0$ nm. Likewise, $EL_{TE1}$ for MFS = 100 nm in SWG ($EL_{TE1}^{SWG,100}$) and conventional ($EL_{TE1}^{Conv,100}$) Y–junctions are depicted in Fig. 8(b). As previously mentioned, $EL_{TE1}^{Conv,50}$ and $EL_{TE1}^{Conv,100}$ are negligible due to TE$_1$ mode odd symmetry, with a power minimum at the center of the multimode stem. Our device exhibits an $EL_{TE1}^{SWG,50}$ below 1 dB over a 100 nm bandwidth ranging from 1475 to 1575 nm. Within a 170 nm bandwidth (1420 – 1590 nm), $EL_{TE1}^{SWG,50}$ and $EL_{TE1}^{SWG,100}$ only increase by 0.5 dB. Compared to the conventional Y–junction, the degradation of the TE$_1$ mode for our SWG Y–junction arises from the selection of the stem waveguide width of $W_S$ = 1200 nm, as a compromise between $EL_{TE0}$ and $EL_{TE1}$. Note that increasing the width of the SWG stem waveguide results in a stronger modal confinement that prevents TE$_1$ mode radiation, but increasingly penalizes TE$_0$ due to the resulting mode profile at the junction tip [57]. Nevertheless, the performance of the proposed device compares very favorably to state-of-the-art higher-order mode power splitters (see Table 1).

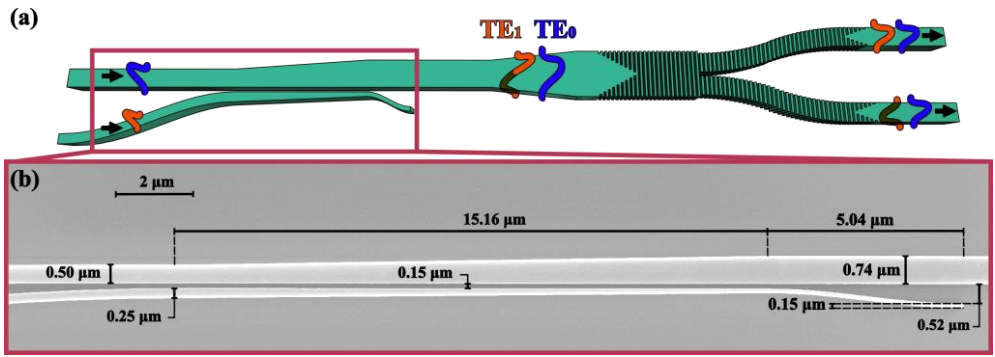

**Fig. 7.** Schematic of the structure employed for the characterization of TE$_1$ mode (a) and SEM images of the mode multiplexer (b).

Table 1. Experimental performance comparison of state-of-the-art multimode power dividers.

| Ref. | Design method | EL (dB) | Bandwidth (nm) | MFS (nm) | Length (µm) | Functionality |
|---|---|---|---|---|---|---|
| [33] | Inverse design subwavelength axisymmetric | 1.5 | 60 | 30 | 2.88 | 2-mode splitter |
| [34] | Pixelated meta-structure | 1.5 | 40 | 30 | 4.5 | 3-mode splitter |
| [60] | MMI coupler | 0.76 | 60 | 1000 | 86.5 | 2-mode splitter |
| [61] | Tapered directional coupler | 0.7 | 30 | 200 | 25 | 2-mode splitter |
| [62] | Densely-packed waveguide array | 1 | 28 | 110 | 46 | 2-mode splitter |
| This work | SWG Y–junction | 1 | 100 | 50 | 41.3 | 2-mode splitter & converter |
| This work | SWG Y–junction | 1.5 | 170 | 100 | 41.3 | 2-mode splitter & converter |

SWG Y–junctions with $\Delta\delta = \pm 10$ nm were also included in combination with the mode multiplexer structures to study fabrication tolerances for TE$_0$ and TE$_1$ mode division mux/demux and are shown in Figs. 9(a) and (b). Figure 9(a) shows that $EL_{TE0}^{SWG,50}$ is lower than 0.5 dB for both under-etching and over-etching errors in the full 1420 nm – 1680 nm bandwidth. Figure 9(b) shows that $EL_{TE0}^{SWG,100}$ remain <1 dB for $\Delta\delta$ = -10 nm and <0.7 dB for $\Delta\delta$ = +10 nm, for the same bandwidth. These results corroborate the robustness of the SWG Y–junction to fabrication errors for TE$_0$ mode, shown in section 3.1 (see Figs. 6(a) and (b)). The performance for TE$_1$ mode division is more sensitive to fabrication errors in the width of the SWG stem owing to the weaker confinement of the Bloch-Floquet TE$_1$ mode compared to the Bloch-Floquet TE$_0$ mode. While under-etching results in $EL_{TE1}^{SWG,50}$ < 2.1 dB and $EL_{TE1}^{SWG,100}$ < 1.7 dB for the full bandwidth, over-etching errors are negligible at shorter wavelengths and increase towards longer wavelengths.

## 4. Conclusions

The detailed experimental study conducted in this work demonstrates the broadband performance and relaxed fabrication tolerances of SWG-based Y–junctions. Two resolutions were investigated to account for different fabrication processes, namely MFS of 50 nm and 100 nm. Accurate measurements in cascaded splitters demonstrate excess loss for the fundamental TE mode lower than 0.3 dB for MFS = 50 nm and below 0.5 dB for MFS = 100 nm, in a 260 nm bandwidth (1420 nm – 1680 nm). Characterization of the first-order TE mode was performed in combination with a mode multiplexer, showing excess loss lower than 1 dB over a 100 nm bandwidth (1475 nm – 1575 nm) for the 50 nm MFS. SWG Y–junctions with deterministically induced errors of -10 nm and +10 nm were also measured to analyze resilience to over- and under-etching errors. Experimental results demonstrate robust fabrication tolerances for the fundamental TE mode. Our SWG-engineered metamaterial Y–junction opens up promising prospects for improving performance of diverse silicon photonic integrated circuits where power splitters are ubiquitous component, such as on-chip high bandwidth communication systems and broadband spectroscopic systems.

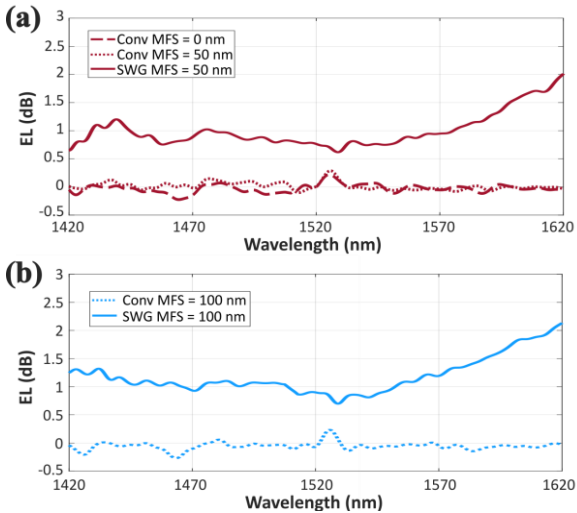

**Fig. 8.** $EL_{TE1}$ measurements for SWG (solid) and conventional (dotted) Y–junctions for MFS of 50 nm (a) and 100 nm (b). Conventional Y–junction (dashed) with MFS = 0 nm is also shown in panel (a).

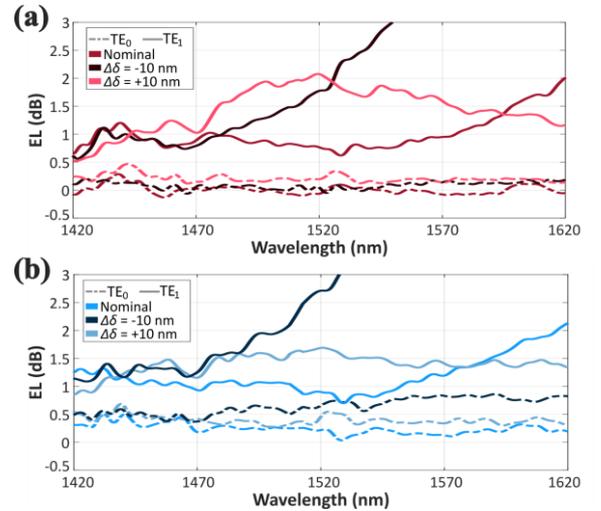

**Fig. 9.** Tolerances to fabrication errors of the SWG Y–junction for TE$_0$ (dashed line) and TE$_1$ (solid line) polarizations, for MFS of 50 nm (a) and 100 nm (b).

**Credit authorship contribution statement**

**Raquel Fernández de Cabo:** Software, Formal Analysis, Validation, Data curation, Writing – original draft. **Jaime Vilas:** Experimental validation, Writing – review & editing. **Pavel Cheben:** Conceptualization, Writing – review & editing. **Aitor V. Velasco:** Conceptualization, Resources, Funding acquisition, Supervision, Writing – review & editing. **David González-Andrade:** Methodology, Supervision, Project administration, Writing – review & editing.

**Declaration of Competing Interest**

The authors declare that they have no known competing financial interests or personal relationships that could have appeared to influence the work reported in this paper.

**Acknowledgements**

This work has been funded by the Spanish Ministry of Science and Innovation (RTI2018-097957-B-C33, RED2018-102768-T, PID2020-115353RA-I00); the Spanish State Research Agency and the European Social Fund Plus under grant PRE2021-096954; the Community of Madrid – FEDER funds (S2018/NMT-4326); the Horizon 2020 research and innovation program under Marie Sklodowska-Curie grant No. 101062518; European Union – NextGenerationEU, through the Recovery, Transformation and Resilience Plan (DIN2020-011488).